\documentclass[a4paper,10pt]{article}
\usepackage{amsfonts}
\usepackage{amsmath}
\usepackage{amssymb}
\usepackage{slashed}
\usepackage[dvips]{graphicx}

\DeclareGraphicsExtensions{.bmp,.png,.pdf,.jpg}

\begin{document}

\title{Entanglement entropy between virtual and real excitations in quantum
electrodynamics}
\author{Juan Sebasti\'{a}n Ardenghi$^{\dag \ddag }$\thanks{%
email:\ jsardenghi@gmail.com, fax number:\ +54-291-4595142} \\
$^{\dag }$Departamento de F\'{\i}sica, Universidad Nacional del Sur, Av.
Alem 1253,\\
B8000CPB, Bah\'{\i}a Blanca, Argentina \\
$^{\ddag }$Instituto de F\'{\i}sica del Sur (IFISUR, UNS-CONICET), Av. Alem
1253,\\
B8000CPB, Bah\'{\i}a Blanca, Argentina}
\maketitle

\begin{abstract}
The aim of this work is to introduce the entanglement entropy of real and
virtual excitations of fermion and photon fields. By rewriting the
generating functional of quantum electrodynamics theory as an inner product
between quantum operators, it is possible to obtain quantum density
operators representing the propagation of real and virtual particles. These
operators are partial traces, where the degrees of freedom traced out are
unobserved excitations. Then the Von Neumann definition of entropy can be
applied to these quantum operators and in particular, for the partial traces
taken over the internal or external degrees of freedom. A universal behavior
is obtained for the entanglement entropy for different quantum fields at
zero order in the coupling constant. In order to obtain numerical results at
different orders in the perturbation expansion, the Bloch-Nordsieck model is
considered, where it it shown that for some particular values of the
electric charge, the von Neumann entropy increases or decreases with respect
to the non-interacting case.
\end{abstract}

\section{Introduction}

Entanglement entropy has become an important topic in theoretical physics
and has become a widely studied topic in the last few years. In general, the
entanglement is studied between one part of a system and in different
branches of theoretical physics usually the partitioning is spatial. An
entanglement entropy can be defined through the quantum density operator and
permits applying the concept in different frameworks, for example to
distinguish new topological phases and characterize critical points (\cite%
{levin}, \cite{kita} and \cite{hsu}) or in discussions of holographic
descriptions of quantum gravity, in particular, for the AdS/CFT
correspondence (\cite{ryu}). More recently the entanglement entropy has been
applied in condensed matter physics, density matrix renormalization group
method (\cite{vidal}, \cite{oster}) and black hole thermodynamics (see \cite%
{Met}, \cite{nis}, \cite{casini}, \cite{ryu}, \cite{solo} and \cite{fur}),
thermal quantum field theory (see \cite{wolf}, \cite{gio} and \cite{cramer})
curved space-time (see \cite{solo2}, \cite{fur3} and \cite{ichi}),
decoherence \cite{lombar}, squeezed vacuum \cite{bianchi2} and in low
dimension systems \cite{bianchi3}.

The concept of entanglement entropy in quantum field theory is linked to a
region of space-time that contains the relevant degrees of freedom (\cite%
{iss}, \cite{nis2}, \cite{amin} and \cite{niza}). The trace over the degrees
of freedom localized on a region which is not accessible to the observer,
results in a reduced density matrix. Then, the von Neumann definition of
entanglement entropy can be applied to obtain a measure of the
inaccesibility of the vacuum state that is mixed after the partial trace. In
QFT geometric entropy can be computed by using the Euclidean path integral
method in models without interactions and the results show that in $d$
dimensions, the entropy behaves as a Laurent series starting in $\epsilon
^{-(d-1)}$, where $\epsilon $ is a short-distance cutoff and the leading
coefficient that multiplies to $\epsilon ^{-(d-1)}$ is proportional to the $%
d-1$ power of the size of $V$, which is the area law for the entanglement
entropy \cite{casini}.

Although entanglement entropy in quantum field theory has been focused on
entanglement between degrees of freedom associated with spatial regions, it
is also permissible to consider the entanglement between real and virtual
excitations. The virtual excitations are a mere mathematical artifact of the
pertubation expansion, so in principle any physical quantity that depends on
this entanglement depends naturally on interactions introduced in the
Lagrangian. On the other hand, given that interactions introduce virtual
excitations and these are entangled with the real excitations, then an
interaction entanglement entropy can be defined and it would be a measure of
the information restored in the propagation of the quantum field, this
information would depend on the interactions with other quantum fields or
itself. In \cite{jsa}, the generating functional of the $\phi ^{4}$ theory
has been written in terms of quantum operators. These operators are partial
traces over larger quantum operators that depends on the internal vertices
and a new set of vertices. These new vertices imply that there are real
particles propagating elsewhere but cannot be measured; then we must average
over the possible space-time points where these particles propagate. This
inaccesibility to these new particles implies that there are unobserved
particles or virtual particles. That is, interactions introduce new
particles, but these particles cannot be observed, then the quantum state
must be traced out. Because the real particles and the new particles are
entangled, then the entanglement entropy can be computed. A very simple
example (see \cite{jsa}) is the first order correction to the $\phi ^{4}$
theory, where the quantum density operator can be written as (not normalized)%
\begin{equation}
\rho =\int \Delta (x_{1}-y_{1})\Delta (y_{1}-w_{1})\Delta
(y_{1}-x_{2})\left\vert x_{1},y_{1}\right\rangle \left\langle
x_{2},w_{1}\right\vert d^{4}w_{1}d^{4}y_{1}d^{4}x_{1}d^{4}x_{2}  \label{q1}
\end{equation}%
By considering the following quantum operator 
\begin{equation}
O=\int \delta (y_{1}-w_{1})J(x_{1})J(x_{2})\left\vert
x_{1},y_{1}\right\rangle \left\langle x_{2},w_{1}\right\vert
d^{4}w_{1}d^{4}y_{1}d^{4}x_{1}d^{4}x_{2}  \label{q2}
\end{equation}%
where $J(x)$ are the external sources and the Dirac delta $\delta
(y_{1}-w_{1})$ is explicitly shown inside the integral in order to remark
that the coefficients related to the internal degrees of freedom are the
identity matrix. Then it follows that the mean value $Tr(\rho O)$ is
identical to the first order in $\lambda _{0}$ of the generating functional.
In turn, $Tr(\rho O)=Tr(\rho _{ext}O_{ext})$, where 
\begin{equation}
\rho _{ext}=Tr_{int}(\rho )=\Delta (0)\int \Delta (x_{1}-y_{1})\Delta
(y_{1}-x_{2})d^{4}y_{1}\left\vert x_{1}\right\rangle \left\langle
x_{2}\right\vert d^{4}x_{1}d^{4}x_{2}  \label{q3}
\end{equation}%
which is identical to the first correction to the two-point correlation
function. The trace over the internal degrees of freedom $y_{1}$ and $w_{1}$
implies that there is a virtual propagation between $y_{1}$ and $w_{1}$ that
is unobserved and then their degrees of freedom must be traced out. This is
the crucial point of the idea of this manuscript and \cite{jsa}: the quantum
operator of the quantum field theory is a partial trace which implies, in
some sense, that some physical process has been neglected and moreover, the
consequences of this lack of observability occurs in the scattering
processes of $\phi ^{4}$ theory\textit{. }The coefficient of the quantum
density operator $\rho $ is entangled in the coordinates because these are
linked through the propagators. Making a Fourier transform, the quantum
operator can be written in the momentum basis as%
\begin{equation}
\rho =\int \int \frac{d^{D}p}{(2\pi )^{D}}\frac{d^{D}q}{(2\pi )^{D}}\frac{%
d^{D}r}{(2\pi )^{D}}\frac{1}{p_{1}^{2}-m_{0}^{2}}\frac{1}{p_{2}^{2}-m_{0}^{2}%
}\frac{1}{p_{3}^{2}-m_{0}^{2}}\left\vert
p_{1},p_{2}+p_{3}-p_{1}\right\rangle \left\langle p_{3},p_{2}\right\vert
\label{q4}
\end{equation}%
In this way, the coefficient is not entangled, each propagator depends on
its momentum vector but the entanglement has been translated to the bra and
ket vectors. That is, the degrees of freedom of an interacting quantum field
theory are entangled in momentum space \cite{bala}.

In \cite{PR1}, \cite{PR2} and \cite{PR3}, the full description of the model
described above is done, where the intermediate operators introduced
artificially by the perturbation expansion can be obtained as partial traces
over the internal degrees of freedom, represented by a duplication of the
internal vertices of the internal propagators. The particles that are
created in these vertices are virtual particles because they do not obey the
constraint of the energy-momentum relation. This implies that these
particles are not measured, then it must be traced out. This unobservation
causes these particles to become virtual. One of the most known consequence
of the imposibility of unobserved particles is in the scattering process of
quantum electrodynamics (QED), where the infrared divergences are canceled
by the contribution of the soft photons which are unobserved photons (\cite%
{kino} and \cite{lee}). Although this phenomena will be discussed in the
next section in relation with the photon entropy, it must be stressed that
the additional soft photon emmisions can be interpreted as "opened" loops in
the scheme presented in \cite{jsa} (figure 1). It should be stressed that in
the previous work \cite{jsa}, the quantum operators $\rho $ and $O$ has been
called "states" and "observables". Although the main result of this work,
where the correlation function can be written as $Tr(\rho O)$ suggests to
consider $\rho $ as a quantum state written formally as a quantum density
operator and $O$ as an observable, the mathematical objects cannot be
associated to physical concepts, mainly because the latter can be
constrained by physical relations, where the former are defined
mathematically. In particular, the quantum states satisfy dynamical
equations and the quantum density operators $\rho $ defined in \cite{jsa}
using the generating functional obeys a functional differential equation
(see eq.(1) of page 288 of \cite{greiner}). In this sense, the quantum
entropy computed can be related to processes, but not to quantum states.

The model introduced in this work can be considered a particular case of the
General Boundary Formalism (GBF) (\cite{oeckl1}, \cite{oeckl2}, \cite{oeckl3}%
, \cite{oeckl4} and \cite{oeckl5}), where to each boundary $S$ defined by an
space-like hyperplane in Minskowski space-time there is a vector space $%
\mathcal{H}_{s}$. In turn, for a given boundary $S$ changing the orientation
corresponds to replace $\mathcal{H}_{s}$ with $\mathcal{H}_{s}^{\ast }$.
Moreover, associated with $M$, which is the region bounded by $S$, there is
a complex function $\rho _{M}:\mathcal{H}_{s}\rightarrow 
\mathbb{C}
$ which associates an amplitude to a state. In turn, if $S$ can be
decomposed into disconnected componentes $S=S_{1}\cup S_{2}...\cup S_{n}$,
then one may convert $\rho _{M}:\mathcal{H}_{S_{1}}\otimes ...\otimes 
\mathcal{H}_{S_{n}}\rightarrow 
\mathbb{C}
$ to a function $\rho _{M}:\mathcal{H}_{S_{1}}\otimes ...\otimes \mathcal{H}%
_{S_{k}}\rightarrow \mathcal{H}_{S_{k+1}}^{\ast }\otimes ...\otimes \mathcal{%
H}_{S_{n}}^{\ast }$ replacing spaces with dual spaces. In the general
boundary formalism, then the focus is moved from quantum states, which
describe a system at some given time, to quantum states of processes, which
describe what happens to a local system during a finite time-span. For
conventional nonrelativistic system, the quantum space of the processes are
defined as the tensor product of the initial and final Hilbert state spaces $%
\mathcal{H}_{1}\otimes \mathcal{H}_{2}$ where the subscripts $1$ and $2$
indicate the initial and final stages of the process. The amplitude of the
process is represented by the Feynman propagator and is determined as a
linear functional over the quantum state defined as the tensor product of
the initial and final quantum states.\footnote{%
In Appendix A a closer relation between the General Boundary Formalism and
the model introduced in this work is discussed.} In \cite{jsa}, the
processes are ordered in terms of the perturbation parameter $\lambda _{0}$.
The external points of the correlation functions define the boundary and
this boundary should be chosen as space-like hyperplanes as it is done in 
\cite{oeckl4}, which implies to fix the time components of $x_{1}$ and $%
x_{2} $ and consider $\mathcal{H}_{1}\otimes \mathcal{H}_{2}$ as the space
which represents the whole family of transition amplitudes between two
space-like hyperplanes.\footnote{%
In the general boundary formalism, the observables defined in the
preparation stage are written as $O\otimes I$, where the identity acts on
the bulk and in the measurement stage the observable is written as $I\otimes
O$. This is similar of what happens in the observable-state model, where an
identity in the observables implies to traced out the irrelevant degrees of
freedom that appears in the perturbation expansion. That is, interactions
introduce new sets of Hilbert spaces, but the observables defined on it
contain identity operators. Then it appears that self-interactions in
quantum scalar fields can be related to the quantum states of the bulk of
the boundaries. The utility of the observable-state model is that the
complexity of the Hilbert space structure depends on the order of the
perturbation expansion.} But when interactions are turned on, internal
propagators appear and moreover, we must integrate over the possible
space-time coordinate of these propagators. Must be stressed that to
integrate in the external points, implies to connect $x_{1}$ with $x_{2}$ in
the Feynman diagrams, which is a simple example of the generation of
correlation functions from vacuum diagrams (see Section 5.5 of \cite%
{kleinert}, page 68), where for example, by cutting one line to the first
order vacuum diagram we obtain the first order contribution to the two-point
function. Then, the space-time coordinates should not be fixed when
interactions are considered because all the correlation functions are
related. A vertex inside a Feynman diagram can be converted into an external
point by cutting an internal propagator (see \cite{jsa}). As an example of
the concept of family of processes, we can consider the Feynman propagator
in 3+1 dimensions (see \cite{chin}) 
\begin{gather}
\Delta (s)=\theta (s^{2})\frac{im}{8\pi \sqrt{s^{2}-i\epsilon }}H_{1}^{(2)}(m%
\sqrt{s^{2}-i\epsilon })+  \label{op1} \\
\theta (-s^{2})\frac{im}{8\pi \sqrt{-s^{2}+i\epsilon }}K_{1}(m\sqrt{%
-s^{2}+i\epsilon })  \notag
\end{gather}%
where $s^{2}=\Delta t^{2}-\Delta r^{2}$ is the proper distance and $%
H_{1}^{(2)}$ is the Hankel function of the second kind and $K_{1}$ is the
modified Bessel function of the first kind. What is interesting of this
propagator is that depends on the proper distance between the two space-time
coordinates. We can consider the whole family of processes that is
parametrized by $s$. For $s\in (-\infty ,0)$ \ we have spacelike interval, $%
s=0$ lightlike interval and $s\in (0,\infty )$ timelike interval. We can
consider that we are only interested in those processes with timelike
interval, then if we consider $\Delta (s)$ the amplitude of the process,
then $\left\vert \Delta (s)\right\vert ^{2}$ is the probability of the
process. If we demand that it is a probability then it must be normalized,
which can be obtained easily for timelike intervals $\int_{0}^{\infty
}\left\vert \Delta ^{2}(s)\right\vert ds=\frac{m^{3}}{2\pi ^{3}}$ \footnote{%
Section 6 of \cite{int} was used to compute the integrals.}.

In \cite{bianchi}, the distinction between pure and mixed states is weaken
in the general covariant context when finite spatial regions are considered.
In the model introduced in this paper, the quantum state is mixed when
interactions are turned on. The mixture is due to the entanglement of the
virtual state in the bulk with the real states in the boundary. In turn, for
free fields there is a priori distinction between pure and mixture states
because we can distinguish between past and future parts of the boundary.
Moreover, the observables acts in the infinite past and infinite future. In
this sense, it seems that the model introduced in the manuscript submitted
is a particular case of the general boundary formalism with the
incorporation of the interactions treated in a perturbative manner and
allowing these virtual states to be defined in the whole space-time.

In order to introduce the formalism for quantum operators and where the
trace can be applied, the generating functional of the quantum field theory
must be considered. As it was done for the self-interacting theory $\phi
^{4} $, it is necessary to establish the formalism to the quantum field
theory of electrons, positrons and photons in order to apply the concept of
entanglement entropy between these particles. Due to the complicated
integrals that must be solved, in order to obtain results for the second
order corrections to the photonic and fermionic entropies, the
Bloch-Nordsieck model \cite{Bloch} will be considered, to show the way in
which the interaction entanglement introduces changes in the quantum
entropy. Then, the manuscript will be organized as follows: In Section II,
the formalism for the quantum opearators by rewriting the generating
functional is introduced for electrons and photons. In section III, the von
Neumann entropy is computed for the electron and photon propagator at zero
order in $e$ and the first corrections are sketched by using the results
found in Appendix B. The Bloch-Nordsieck model is discussed and exact
results for the von Neumann entropy are obtained. In last section, the
conclusions are presented and in Appendix A a conceptual discussion of the
model is done.

\section{Quantum operators in QED}

\begin{figure}[tbp]
\centering\includegraphics[width=125mm,height=80mm]{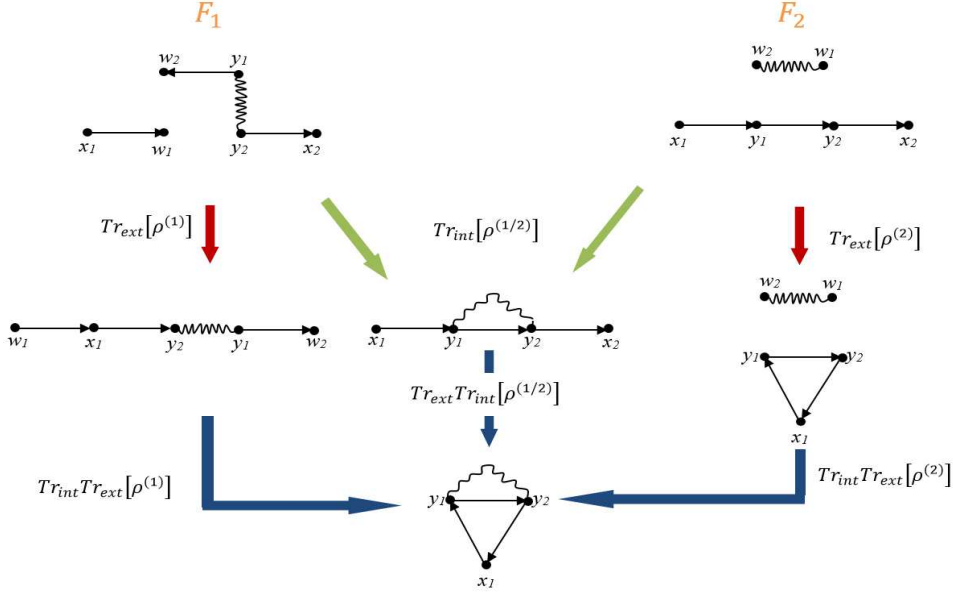}
\caption{Scheme of partial traces over the possible quantum operator of the
electron propagator at first order in $e$. }
\label{diagrams}
\end{figure}

The generating functional can be constructed in a general way by considering
some (symmetric) $n$-point functions $\tau ^{(n)}(x_{1},...,x_{n})$, then
the corresponding generating functional (\cite{Haag}, eq. (II.2.21), \cite%
{Brown}, eq. (3.2.11)) can be defined as%
\begin{equation}
Z\left[ \eta ,\overline{\eta }\right] =\underset{n=0}{\overset{\infty }{\sum 
}}\frac{i^{n}}{n!}\int \tau ^{(n)}(x_{1},...,x_{n})\eta (x_{1})\overline{%
\eta }(x_{2})...\eta (x_{n-1})\overline{\eta }(x_{n})\underset{i=1}{\overset{%
n}{\prod }}d^{4}x_{i}  \label{ideas1}
\end{equation}%
where $\eta (x_{i})$ and $\overline{\eta }(x_{i})$ are external sources for\ 
$\psi (x_{i})$ and $\overline{\psi }(x_{i})$ fields respectively and $\tau
^{(n)}$ can be%
\begin{eqnarray}
\tau _{F}^{(n)}(x_{1},...,x_{n}) &=&S^{(n)}(x_{1},...,x_{n})=\left\langle
\Omega \left\vert \psi (x_{1})\overline{\psi }(x_{2})...\psi (x_{n-1})%
\overline{\psi }(x_{n})\right\vert \Omega \right\rangle  \label{ideas2} \\
\tau _{P}^{(n)}(x_{1},...,x_{n}) &=&D_{\mu _{1},...,\mu
_{n}}^{(n)}(x_{1},...,x_{n})=\left\langle \Omega \left\vert A_{\mu
_{1}}(x_{1})...A_{\mu _{n}}(x_{n})\right\vert \Omega \right\rangle  \notag
\end{eqnarray}%
where the first correlation function is for fermions and the second is for
photons, $\psi (x)$ ($\overline{\psi }(x)$) and $A_{\mu }(x)$ are the
fermion (positron) and photon fields and $\left\vert \Omega \right\rangle $
is the vacuum state. A convenient way to eliminate trivial contributions in
the correlation function is by introducing a modified generating functional $%
Z\left[ \eta ,\overline{\eta }\right] $ for irreducible Green's functions
that is defined as $W\left[ \eta ,\overline{\eta }\right] =e^{iZ\left[ \eta ,%
\overline{\eta }\right] }$. The new generating functional $Z\left[ \eta ,%
\overline{\eta }\right] $ satisfies the normalization condition $Z[0,0]=0$
and it reads%
\begin{equation}
Z\left[ \eta ,\overline{\eta }\right] =\underset{n=0}{\overset{\infty }{\sum 
}}\frac{1}{n!}\int \tau _{c}^{(n)}(x_{1},...,x_{n})\eta (x_{1})\overline{%
\eta }(x_{2})...\eta (x_{n-1})\overline{\eta }(x_{n})\underset{i=1}{\overset{%
n}{\prod }}d^{4}x_{i}  \label{ideas4}
\end{equation}%
where in this case $\tau _{c}^{(n)}(x_{1},...,x_{n})$ are connected $n$%
-point functions that can be obtained by differentiation%
\begin{equation}
\tau _{c}^{(n)}(x_{1},...,x_{n})=\frac{\delta ^{n}Z\left[ \eta ,\overline{%
\eta }\right] }{\delta \eta (x_{1})\delta \overline{\eta }(x_{2})...\delta
\eta (x_{n-1})\delta \overline{\eta }(x_{n})}\mid _{\substack{ \eta =0  \\ 
\overline{\eta }=0}}  \label{ideas5}
\end{equation}%
In turn, the connected $n$-point functions can be written in terms of the
Lagrangian interaction density $\mathcal{L}_{I}^{0}=-e\overline{\psi }\gamma
^{\mu }\psi A_{\mu }$ for QED as (see eq.(II.2.33) of \cite{Haag})\footnote{%
In eq.(\ref{ideas6}) we have introduced the perturbative expansion of the
correlation function, where the $y_{i}$ are the internal vertices.}%
\begin{equation}
S^{(n)}(x_{1},...,x_{n})^{(p)}=\frac{i^{p}}{p!}\int \left\langle \Omega
_{0}\left\vert T\psi (x_{1})\overline{\psi }(x_{2})...\psi (x_{n-1})%
\overline{\psi }(x_{n})\mathcal{L}_{I}^{0}(y_{1})...\mathcal{L}%
_{I}^{0}(y_{p})\right\vert \Omega _{0}\right\rangle \underset{i=1}{\overset{p%
}{\prod }}d^{4}y_{i}  \label{ideas6}
\end{equation}%
for external fermions and%
\begin{equation}
D_{\mu _{1},...,\mu _{n}}^{(n)}(x_{1},...,x_{n})^{(p)}=\frac{i^{p}}{p!}\int
\left\langle \Omega _{0}\left\vert TA_{\mu _{1}}(x_{1})...A_{\mu _{n}}(x_{n})%
\mathcal{L}_{I}^{0}(y_{1})...\mathcal{L}_{I}^{0}(y_{p})\right\vert \Omega
_{0}\right\rangle \underset{i=1}{\overset{p}{\prod }}d^{4}y_{i}
\label{ideas6.0}
\end{equation}%
for external photons and introducing (\ref{ideas6}) in (\ref{ideas4}) we have%
\begin{gather}
iZ_{F}\left[ \eta ,\overline{\eta }\right] =\underset{n=0}{\overset{\infty }{%
\sum }}\underset{p=0}{\overset{\infty }{\sum }}\frac{i^{n}}{n!}\frac{i^{p}}{%
p!}\int \left\langle \Omega _{0}\left\vert T\psi (x_{1})\overline{\psi }%
(x_{2})...\psi (x_{n-1})\overline{\psi }(x_{n})\mathcal{L}_{I}^{0}(y_{1})...%
\mathcal{L}_{I}^{0}(y_{p})\right\vert \Omega _{0}\right\rangle
\label{ideas7} \\
\eta (x_{1})\overline{\eta }(x_{2})...\eta (x_{n-1})\overline{\eta }(x_{n})%
\underset{i=1}{\overset{n}{\prod }}d^{4}x_{i}\underset{i=1}{\overset{p}{%
\prod }}d^{4}y_{i}  \notag
\end{gather}%
and%
\begin{gather}
iZ_{P}\left[ \eta ,\overline{\eta }\right] =\underset{n=0}{\overset{\infty }{%
\sum }}\underset{p=0}{\overset{\infty }{\sum }}\frac{i^{n}}{n!}\frac{i^{p}}{%
p!}\int \left\langle \Omega _{0}\left\vert TA_{\mu _{1}}(x_{1})...A_{\mu
_{n}}(x_{n})\mathcal{L}_{I}^{0}(y_{1})...\mathcal{L}_{I}^{0}(y_{p})\right%
\vert \Omega _{0}\right\rangle \eta (x_{1})  \label{ideas7.1} \\
\overline{\eta }(x_{2})...\eta (x_{n-1})\overline{\eta }(x_{n})\underset{i=1}%
{\overset{n}{\prod }}d^{4}x_{i}\underset{i=1}{\overset{p}{\prod }}d^{4}y_{i}
\notag
\end{gather}%
where in last equation, indices in $Z_{p}\left[ \eta ,\overline{\eta }\right]
$ are not written. The main idea on which is based the entanglement entropy
between real and virtual field excitations is that both generating
functionals can be written as an inner product of a quantum operator defined
through the $\eta (x)$ and $\overline{\eta }(x)$ sources with a quantum
operator defined by the correlation functions $S^{(n)}(x_{1},...,x_{n})$ and 
$D_{\mu _{1},...,\mu _{n}}^{(n)}(x_{1},...,x_{n})$. For the sake of
simplicity, the procedure will be shown for the generating functional of the
fermion correlation functions. The procedure for photon correlation
functions is identical. To define the quantum operator we can consider some
operator function $\mathbf{F}$ that depends on a set of vertices $y_{1}$,...,%
$y_{p}$ and some new coodinates $w_{1},...,w_{p}$ in such a way that%
\begin{equation}
\int \mathbf{F}(y_{1},..,y_{p},w_{1},...,w_{p})\underset{i=1}{\overset{p}{%
\prod }}\delta (y_{i}-w_{i})d^{4}w_{i}=\mathcal{L}_{I}^{0}(y_{1})...\mathcal{%
L}_{I}^{0}(y_{p})  \label{os1}
\end{equation}%
where $\mathcal{L}_{I}^{0}(y_{p})$ is the Lagrangian that appears in eq.(\ref%
{ideas6}). In \cite{PR1} we have studied the $\phi ^{4}$ theory and two
possible functional forms can be found. In a similar way, the corresponding
operator for quantum electrodynamics $\mathcal{L}_{I}^{0}=-e\overline{\psi }%
\gamma ^{\mu }\psi A_{\mu }$ can be represented by two different functional
forms%
\begin{eqnarray}
\mathbf{F_{1}}(y_{1},..,y_{p},w_{1},...,w_{p}) &=&(-1)^{p}e^{p}\underset{i=1}%
{\overset{p}{\prod }}\overline{\psi }(y_{i})\gamma ^{\mu _{i}}\psi
(w_{i})A_{\mu _{i}}(y_{i})  \label{os2} \\
\mathbf{F_{2}}(y_{1},..,y_{p},w_{1},...,w_{p}) &=&(-1)^{p}e^{p}\underset{i=1}%
{\overset{p}{\prod }}\overline{\psi }(y_{i})\gamma ^{\mu _{i}}\psi
(y_{i})A_{\mu _{i}}(w_{i})  \notag
\end{eqnarray}%
where in both cases, eq.(\ref{os1}) holds, that is, by introducing $\mathbf{F%
}_{1/2}$ in eq.(\ref{os1}), and performing the integration in $w_{1}$ using
the the Dirac delta $\delta (y_{i}-w_{i})$, the QED Lagrangian is recovered $%
\mathcal{L}_{I}^{0}(y)=$ $e\overline{\psi }(y)\gamma ^{\mu }\psi (y)A_{\mu
}(y)$. The main difference between $\mathbf{F_{1}}$ and $\mathbf{F_{2}}$ is
that the new internal vertex $w_{i}$ is attached to the fermion field for $%
\mathbf{F_{1}}$ and to the photon field for $\mathbf{F_{2}}$.\textit{\ }The
last equation implies that are we are considering a non-local Lagrangian
that contains information that can be traced out. It must be stressed that
although there are two different ways to introduce the formalism, for the
purposes of this work, any choice would be adequate because, as was shown in
eq.(\ref{os1}), the quantum operator that appears in the correlation
function of QED is the reduced quantum operator, which does not depend on
the prescription adopted $\mathbf{F_{1}}$ or $\mathbf{F_{2}}$. Different von
Neumann entropies will be obtained for the non-traced quantum operator
whereas for the reduced operators the von Neumann entropy is identical for
both prescriptions (see figure \ref{diagrams}). In \cite{jsa} a physical
interpretation of the operator function $\mathbf{F_{i}}$ is given for $\phi
^{4}$ theory. In the same way, we can consider $\mathbf{F_{1}}$ in eq.(\ref%
{os2}) for the electron propagator.\footnote{%
The same partition can be found in \cite{jsa}. A way to explain both
partitions is by considering the quantum operator defined as 
\begin{equation*}
\rho _{1}=\left\vert \phi _{0}(x_{1})\right\rangle \left\langle \phi
_{0}(x_{2})\right\vert \otimes \left\vert \phi _{0}^{2}(y_{1})\right\rangle
\left\langle \phi _{0}^{2}(w_{1})\right\vert
\end{equation*}%
for the first partition and 
\begin{equation*}
\rho _{2}=\left\vert \phi _{0}(x_{1})\right\rangle \left\langle \phi
_{0}(x_{2})\right\vert \otimes \left\vert \phi _{0}^{3}(y_{1})\right\rangle
\left\langle \phi _{0}(w_{1})\right\vert
\end{equation*}%
for the second partition in \cite{jsa}. From this point of view, the quantum
operators with interactions can be conceived as composite operator functions.%
} In this case the non-reduced quantum operator represents an electron in a
definite momentum which is prepared in the infinite past $x_{1}$, and when
the interaction is turned on, this electron annihilates at the point $w_{1}$%
. In the point $y_{1}$ an electron and a photon are created, where the
electron annihilates at the point $w_{2}$ and the photon annihilates at the
point $y_{2}$ and creates a new electron that propagates and is measured in
the infinite future point $x_{2}$. In the same way, $\mathbf{F_{2}}$
describes the physical process in which an electron is created at the point $%
x_{1}$ and annihilates at $y_{1}$, where another electron is created and
annihilates at $y_{2}$, where a third electron is created and measured in $%
x_{2}$. At the coordinate $w_{1}$ a photon is created and annihilates at $%
w_{2}$. For experimental purposes, different choices of the operator
function is irrelevant because there is no available experimental procedure
in which the remaining particles propagating elsewhere can be measured in
such a way to to have access to the non-traced quantum state. The unique
comparison available is then the von Neumann entropy with and without
interaction. In order to understand the number of different choices of the
operator function in $\phi ^{n}$ theory, a simple inspection indicates that
a $\phi ^{n}$ theory can be split, according to the partition, to $n=p+q$,
where $p$ and $q$ are natural numbers. Because $n$ is symmetric under
interchange of $p$ and $q$, the number of different splitting is $\frac{n}{2}
$. In \cite{jsa} a particular operator function was adopted because it was
easier to compute the von Neumann entropy of the non-reduced quantum
operator. In this work no prescription is adopted because only the
entanglement entropy of the reduced operator will computed, which does not
depend on the choice of the operator function.

In the first case of eq.(\ref{os2}), in terms of Feynman diagrams, a
positron and a photon field interact at the same space-time point and an
electron in a different point. In the second case, an electron and a
positron interact at the same space-time point and a photon field acts in a
different space-time point. Both functions of the fields will contain the
same reduced state when the internal degrees of freedom are traced out.
Then, inserting eq.(\ref{os1}) in eq.(\ref{ideas7})\ we obtain%
\begin{gather}
iZ_{F}\left[ \eta ,\overline{\eta }\right] =\underset{n=0}{\overset{\infty }{%
\sum }}\underset{p=0}{\overset{\infty }{\sum }}\frac{i^{n}}{n!}\frac{i^{p}}{%
p!}\int \left\langle \Omega _{0}\left\vert T\psi (x_{1})\overline{\psi }%
(x_{2})...\psi (x_{n-1})\overline{\psi }(x_{n})\mathbf{F}%
(y_{1},..,y_{p},w_{1},...,w_{p})\right\vert \Omega _{0}\right\rangle
\label{os4} \\
\eta (x_{1})\overline{\eta }(x_{2})...\eta (x_{n-1})\overline{\eta }(x_{n})%
\underset{i=1}{\overset{p}{\prod }}\delta (y_{i}-w_{i})\underset{i=1}{%
\overset{n}{\prod }}d^{4}x_{i}\underset{i=1}{\overset{p}{\prod }}%
d^{4}y_{i}d^{4}w_{i}  \notag
\end{gather}%
Now we can define two quantum operators in the following way%
\begin{gather}
\varrho ^{(F,n,p)}=\int \left\langle \Omega _{0}\left\vert T\psi (x_{1})%
\overline{\psi }(x_{2})...\psi (x_{n-1})\overline{\psi }(x_{n})\mathbf{F}%
(y_{1},..,y_{p},w_{1},...,w_{p})\right\vert \Omega _{0}\right\rangle
\label{os5} \\
\left\vert x_{1},...,x_{\frac{n}{2}},y_{1},...,y_{p}\right\rangle
\left\langle x_{\frac{n}{2}+1},...,x_{n},w_{1},...,w_{p}\right\vert \underset%
{i=1}{\overset{n}{\prod }}d^{4}x_{i}\underset{i=1}{\overset{p}{\prod }}%
d^{4}y_{i}d^{4}w_{i}  \notag
\end{gather}%
\begin{gather}
O^{(n,p)}=\int \eta (x_{1})\overline{\eta }(x_{2})...\eta (x_{n-1})\overline{%
\eta }(x_{n})\underset{i=1}{\overset{p}{\prod }}\delta
(y_{i}-w_{i})\left\vert x_{1},...,x_{\frac{n}{2}},y_{1},...,y_{p}\right%
\rangle \left\langle x_{\frac{n}{2}+1},...,x_{n},w_{1},...,w_{p}\right\vert
\label{os6} \\
\underset{i=1}{\overset{n}{\prod }}d^{4}x_{i}\underset{i=1}{\overset{p}{%
\prod }}d^{4}y_{i}d^{4}w_{i}  \notag
\end{gather}%
Then, eq.(\ref{os4}) can be written as%
\begin{equation}
iZ_{F}\left[ \eta ,\overline{\eta }\right] =\underset{n=0}{\overset{\infty }{%
\sum }}\underset{p=0}{\overset{\infty }{\sum }}\frac{i^{n}}{n!}\frac{i^{p}}{%
p!}Tr(\varrho ^{(F,n,p)}O^{(n,p)})  \label{os7}
\end{equation}%
The quantum operator of eq.(\ref{os6}) has the following form%
\begin{equation}
O^{(n,p)}=O_{ext}^{(n)}\otimes I_{int}^{(p)}  \label{os8}
\end{equation}%
where%
\begin{equation}
O_{ext}^{(n)}=\int \eta (x_{1})\overline{\eta }(x_{2})...\eta (x_{n-1})%
\overline{\eta }(x_{n})\left\vert x_{1},...,x_{\frac{n}{2}}\right\rangle
\left\langle x_{\frac{n}{2}+1},...,x_{n}\right\vert \underset{i=1}{\overset{n%
}{\prod }}d^{4}x_{i}  \label{os9}
\end{equation}%
and%
\begin{gather}
I_{int}^{(p)}=\int \underset{i=1}{\overset{p}{\prod }}\delta
(y_{i}-w_{i})\left\vert y_{1},...,y_{p}\right\rangle \left\langle
w_{1},...,w_{p}\right\vert \underset{i=1}{\overset{p}{\prod }}%
d^{4}y_{i}d^{4}w_{i}=  \label{os10} \\
\int \left\vert y_{1},...,y_{p}\right\rangle \left\langle
y_{1},...,y_{p}\right\vert \underset{i=1}{\overset{p}{\prod }}d^{4}y_{i} 
\notag
\end{gather}%
is an identity operator acting on the $y_{i}$ vertices that appear in the
perturbation expansion. The Dirac delta that appears as the coefficient of
the identity operator can be considered as a particular choice of an
operator that physically implies no measurement. The subscript $ext$ in eq.(%
\ref{os8}) refers to the external points $x_{i}$ and the subscript $int$ to
the internal vertices $y_{i}$. Then, the generating functional of eq.(\ref%
{ideas7}) can be written as the inner product of the quantum operator $%
O_{ext}$ on the reduced operator $\overline{\varrho }_{ext}$ as%
\begin{equation}
Tr(\varrho ^{(F,n,p)}O^{(n,p)})=Tr(\overline{\varrho }%
_{ext}^{(F,n,p)}O_{ext}^{(n)})  \label{os11}
\end{equation}%
where%
\begin{gather}
\overline{\varrho }_{ext}^{(F,n,p)}=Tr_{int}(\varrho ^{(F,n,p)})=\int
\left\langle y_{1},...,y_{p}\right\vert \varrho ^{(F;n,p)}\left\vert
y_{1},...,y_{p}\right\rangle \underset{i=1}{\overset{p}{\prod }}d^{4}y_{i}=
\label{os12} \\
\int \left( \int\limits_{{}}^{{}}\left\langle \Omega _{0}\left\vert T\psi
(x_{1})\overline{\psi }(x_{2})...\psi (x_{n-1})\overline{\psi }%
(x_{n})L_{I}^{0}(y_{1})...L_{I}^{0}(y_{p})\right\vert \Omega
_{0}\right\rangle \underset{i=1}{\overset{p}{\prod }}d^{4}y_{i}\right) \times
\notag \\
\left\vert x_{1},...,x_{\frac{n}{2}}\right\rangle \left\langle x_{\frac{n}{2}%
+1},...,x_{n}\right\vert \underset{i=1}{\overset{n}{\prod }}d^{4}x_{i} 
\notag
\end{gather}%
The procedure introduced above is suitable to consider the von Neumann
entropy defined as $S_{ext/int}=-Tr\left[ \varrho _{ext/int}\ln (\varrho
_{ext/int})\right] $, where $\varrho _{ext/int}$ are partial traces with
respect the internal/external vertices respectively. In $\phi ^{4}$ theory,
in the propagator, the contributions to the physical mass are given by the
loop diagrams obtained from the perturbation theory. By "opening" the loops,
a quantum density operator can be defined, that represents the propagation
of a defined number of entangled bosons. By considering the internal trace
over this quantum operator, the boson propagator is recovered, represented
by a reduced operator. In this sense, the dressed propagator of the boson is
a reduced operator that represents a real propagating particle entangled
with its virtual excitations and a measure of this entanglement is related
to the physical mass, which is a consequence of the irrelevant degrees of
freedom traced out. In the same way as for the $\phi ^{4}$ theory, we can
write the non-renormalized quantum state of the two-point correlation
function that represents the electron propagation as%
\begin{equation}
\rho _{ext}=\int \frac{d^{4}p}{(2\pi )^{4}}\frac{i}{p^{2}-m_{0}^{2}-\Sigma
(p,m_{0}^{2})}\left\vert p\right\rangle \left\langle p\right\vert
\label{uu1}
\end{equation}%
where $\Sigma (p,m_{0}^{2})$ is the self-energy. For the sake of simplicity,
the first contribution to $\Sigma $ comes from the diagram%
\begin{equation}
\Sigma \sim \lambda _{0}\Delta _{0}+O(\lambda _{0}^{2})\sim \lambda _{0}\int
d^{4}w_{1}\Delta (y_{1}-w_{1})\delta (y_{1}-w_{1})+O(\lambda _{0}^{2})
\label{uu2}
\end{equation}%
Because we can conceive the propagators as quantum density operators, then
it is natural to interpret the coefficients of the operator as the
probability amplitude attached to a particle travelling from one point $%
x_{1} $ to another point $x_{2}$ with an specific value of energy and
momentum that a particle is created at $x_{1}$ and annihilated at $x_{2}$.
Finally, it should be possible to apply the concept of entanglement entropy
between real and virtual excitations for other systems that are treated
perturbatively, for example for the Gell-Mann and Goldberger relation \cite%
{gold1}, in disordered systems in condensed matter \cite{rammer} and
whenever there is a generating functional for the correlation function or a
generating function for the Green functions, as it is occur in condensed
matter with the Luttinger-Ward functional \cite{lut}.

\section{The quantum entropy}

In order to compute the quantum entropy we must take into account the
algebraic structure of the Hilbert space involved in the procedure
introduced in the previous section. The main difference between spinor
quantum electrodynamics and $\phi ^{4}$ theory is that in the latter, the
coefficients of the quantum operators are complex numbers and in the first
theory are $d\times d$ matrices due to the Dirac matrices in $d$ dimensions,
where $d$ is the dimension of space-time when the dimensional regularization
is applied. Nevertheless, the orders of the perturbation considered in this
manuscript implies quantum operators where the $d\times d$ matrices are
identity matrices, then the quantum operators can be written as%
\begin{equation}
\varrho ^{(n)}=[Tr(\varrho ^{(n)})]^{-1}\left[ \varrho ^{(n,0)}\oplus
\varrho ^{(n,1)}\oplus ...\oplus \varrho ^{(n,i)}...\right] =[Tr(\varrho
^{(n)})]^{-1}\underset{j=0}{\overset{+\infty }{\oplus }}\varrho ^{(n,i)}
\label{a1}
\end{equation}%
where the superscript $n~$indicates the number of external points and $i$
indicates the order in the perturbation expansion. $[Tr(\varrho
^{(n)})]^{-1} $ is the normalization of the quantum operator that can be
introduced at the right or left of $\underset{j=0}{\overset{+\infty }{\oplus 
}}\varrho ^{(n,i)} $ because is only a diagonal matrix. The coefficient of
each quantum operator will be of the form%
\begin{gather}
\varrho ^{(n,i)}(x_{1},...,x_{n},y_{1},...,y_{p},w_{1},...,w_{p})=
\label{a2} \\
\left\langle \Omega _{0}\left\vert T\psi (x_{1})\overline{\psi }%
(x_{2})...\psi (x_{n-1})\overline{\psi }%
(x_{n})F(y_{1},..,y_{p},w_{1},...,w_{p})\right\vert \Omega _{0}\right\rangle
\notag
\end{gather}%
The trace reads\footnote{%
Should be clear that the quantum operators $\varrho ^{(n)\text{ }}$that
depend only on the two external points are the partial traces over the
internal degrees of freedom.}%
\begin{equation}
Tr(\varrho ^{(n)})=\underset{j=0}{\overset{+\infty }{\sum }}%
(-e)^{j}W_{(n,j)}Tr(\rho ^{(n,j)})  \label{a3}
\end{equation}%
where $W_{(n,i)}$ is the weight factor (see \cite{PS}) corresponding to the
connected Feynman diagram and $\rho ^{(n,j)}$ is an operator that depends on
the propagator of the respective Feynman diagram. The total quantum entropy
can be computed as%
\begin{equation}
S^{(n)}=-Tr[\varrho ^{(n)}\ln (\varrho ^{(n)})]  \label{a4}
\end{equation}%
where $S$ will be a function of $e$ and some factor which will depend on the
regularization scheme chosen. Up to second order in $e$, the quantum entropy
in terms of $\rho $ reads%
\begin{gather}
S^{(n)}=\ln (\beta ^{(n,0)})-[\beta ^{(n,0)}]^{-1}Tr[\rho ^{(n,0)}\ln (\rho
^{(n,0)})]  \label{a5} \\
-\frac{e^{2}W_{(n,1)}}{W_{(n,0)}[\beta ^{(n,0)}]^{2}}\left[ \beta
^{(n,1)}Tr[\rho ^{(n,0)}\ln (\rho ^{(n,0)})]-\beta ^{(n,0)}Tr[\rho
^{(n,1)}\ln (\rho ^{(n,0)})]\right] +O(\lambda _{0}^{2})  \notag
\end{gather}%
where $\beta ^{(n,i)}=Tr(\rho ^{(n,i)})$.

\subsection{Free fermion field entropy}

In the case of two external points, at zero order in $e$, $\beta
^{(2,0)}=Tr[\rho _{ext}^{(2,0)}]$ and $Tr[\rho _{ext}^{(2,0)}\ln (\rho
_{ext}^{(2,0)})]$ must be computed. The quantum operator at zero order is
the free propagator 
\begin{equation}
\rho _{ext_{F}}^{(2,0)}=\int \frac{d^{4}p}{(2\pi )^{4}}\frac{i(\slashed{p}%
+m_{0})e^{-ip(x_{1}-x_{2})}}{p^{2}-m_{0}^{2}}\left\vert x_{1}\right\rangle
\left\langle x_{2}\right\vert d^{4}x_{1}d^{4}x_{2}  \label{z0}
\end{equation}%
Taking the Fourier transform by writing $\left\vert x_{1}\right\rangle =\int 
\frac{d^{4}q_{1}}{(2\pi )^{4}}e^{-iq_{1}x_{1}}\left\vert q_{1}\right\rangle $
and $\left\langle x_{2}\right\vert =\int \frac{d^{4}q_{2}}{(2\pi )^{4}}%
e^{iq_{2}x_{2}}\left\langle q_{2}\right\vert $, performing a Wick rotation $%
p_{0E}=-ip_{0}$, $p_{iE}=p_{i}$, $d^{4}p=id^{4}p_{E}$ the quantum operator $%
\rho _{ext}^{(2,0)}$ in momentum space is diagonal and reads%
\begin{equation}
\rho _{ext_{F}}^{(2,0)}=\int \frac{d^{4}p_{E}}{(2\pi )^{4}}\frac{(\slashed{p}%
_{E}+m_{0})}{p_{E}^{2}+m_{0}^{2}}\left\vert p_{E}\right\rangle \left\langle
p_{E}\right\vert  \label{z1}
\end{equation}%
where $\slashed{p}_{E}=\gamma _{E}^{\mu }p_{\mu _{E}}$, and $\gamma
_{E}^{\mu }$ are the Euclidean Dirac matrices $\gamma _{E}^{0}=\gamma ^{0}$, 
$\gamma _{E}^{i}=-i\gamma ^{i}$.\footnote{%
A simple inspection implies that $\{\gamma _{E}^{\mu },\gamma _{E}^{\nu
}\}=-2\delta _{\mu \nu }$, $\{\gamma _{E}^{i},\gamma _{E}^{j}\}=-\{\gamma
^{i},\gamma ^{j}\}=2\delta _{ij}I_{4}$, then $(\gamma _{E}^{\mu })^{2}=d$
and\ $\{\slashed{p}_{E},\slashed{q}_{E}\}=2p\cdot q$.} The trace of $\rho
_{ext_{F}}^{(2,0)}$ reads $\beta _{F}^{(2,0)}=Tr[\rho
_{ext_{F}}^{(2,0)}]=2TVm_{0}\Delta _{0}$, where 
\begin{equation}
\Delta _{j}=\int \frac{d^{4}p_{E}}{(2\pi )^{4}}\frac{1}{%
(p_{E}^{2}+m_{0}^{2})^{j+1}}  \label{z2}
\end{equation}%
where the integral of the term with odd $p_{\mu _{E}}$ in the numerator
vanishes by symmetry and where $2TV=\int d^{4}x=\delta ^{(4)}(0)$ (see \cite%
{PS}, page 96). It is interesting to note that $\beta _{F}^{(2,0)}$ for
fermions is different from scalar boson fields, where $\beta
_{B}^{(2,0)}=Tr[\rho _{ext_{B}}^{(2,0)}]=2TV\Delta _{0}$ (see eq.(32) of 
\cite{jsa}).\footnote{%
The extra $i$ factor appears because Wick rotation was not applied.} Because 
$\rho ^{(2,0)}$ is diagonal in the momentum basis, $\ln [\rho
_{ext_{F}}^{(2,0)}]$ reads%
\begin{equation}
\ln [\rho _{ext_{F}}^{(2,0)}]=\int \frac{d^{4}p_{E}}{(2\pi )^{4}}\ln (\frac{%
\slashed{p}_{E}+m_{0}}{p_{E}^{2}+m_{0}^{2}})\left\vert p_{E}\right\rangle
\left\langle p_{E}\right\vert  \label{z3}
\end{equation}%
By computing the matrix logarithm of $\frac{\slashed{p}_{E}+m_{0}}{%
p_{E}^{2}+m_{0}^{2}}$ we obtain (see eq.(\ref{ap5}) of Appendix B)%
\begin{equation}
\ln (\frac{\slashed{p}_{E}+m_{0}}{p_{E}^{2}+m_{0}^{2}})=-\frac{1}{2}\ln
(p_{E}^{2}+m_{0}^{2})I+\frac{\slashed{p}_{E}}{2p_{E}}\ln (\frac{m_{0}+p_{E}}{%
m_{0}-p_{E}})  \label{z3.1}
\end{equation}%
Then, by multiplying eq.(\ref{z3}) with $\rho _{ext}^{(2,0)}$ we obtain%
\begin{equation}
\rho _{ext_{F}}^{(2,0)}\ln [\rho _{ext_{F}}^{(2,0)}]=\int \frac{d^{4}p_{E}}{%
(2\pi )^{4}}\frac{\slashed{p}_{E}+m_{0}}{p_{E}^{2}+m_{0}^{2}}\ln (\frac{%
\slashed{p}_{E}+m_{0}}{p_{E}^{2}+m_{0}^{2}})\left\vert p_{E}\right\rangle
\left\langle p_{E}\right\vert  \label{z3.2}
\end{equation}%
Then the trace $Tr[\rho _{ext_{F}}^{(2,0)}\ln (\rho _{ext_{F}}^{(2,0)})]$
reads%
\begin{equation}
Tr[\rho _{ext_{F}}^{(2,0)}\ln (\rho _{ext_{F}}^{(2,0)})]=TV\left[ \eta
_{0}-m_{0}\chi _{0}\right]  \label{z4}
\end{equation}%
where $\eta _{0}$ reads%
\begin{equation}
\eta _{0}=\int \frac{d^{4}p_{E}}{(2\pi )^{4}}\frac{p_{E}}{p_{E}^{2}+m_{0}^{2}%
}\ln (\frac{m_{0}+p_{E}}{m_{0}-p_{E}})  \label{z5}
\end{equation}%
and%
\begin{equation}
\chi _{0}=\int \frac{d^{4}p_{E}}{(2\pi )^{4}}\frac{\ln (p_{E}^{2}+m_{0}^{2})%
}{p_{E}^{2}+m_{0}^{2}}  \label{z5.1}
\end{equation}%
which has been computed in \cite{jsa}, eq.(36) and eq.(A3) of Appendix B).
In eq.(\ref{z4}) we have disregarded the odd term in $\slashed{p}_{E}$
because it integrates symmetrically to zero. Taking into account all the
terms and using eq.(\ref{a5}) at zero order, the quantum entropy of the free
electron propagation reads%
\begin{equation}
S_{ext_{F}}^{(2)}=\ln (2TVm_{0}\Delta _{0})-\frac{\eta _{0}}{2m_{0}\Delta
_{0}}+\frac{\chi _{0}}{2\Delta _{0}}  \label{z6}
\end{equation}%
where $\Delta _{0}$ was computed in \cite{jsa} using dimensional
regularization. Applying the same regularization scheme in $\eta _{0}$ and $%
\chi _{0}$, the external entropy at zero order in the perturbation expansion
reads 
\begin{equation}
S_{ext_{F}}^{(2)}=-\frac{1}{\epsilon }-\frac{11}{6}+\ln (\frac{m_{0}^{4}TV}{%
4\pi ^{2}\epsilon })+O(\epsilon )  \label{z6.1}
\end{equation}%
where $\epsilon =d-4$ can be considered as a microscopic cutoff. The
appearance of the logarithm of the microscopic cutoff $\epsilon $ has been
obtained in other works \cite{bombelli}, \cite{sred}, \cite{callan}, \cite%
{hol} and \cite{calabre}. The entropy is proportional to the dimensionless
coefficient $\frac{m_{0}^{4}TV}{4\pi ^{2}}$ similar to the result obtained
in \cite{jsa} for the scalar boson which is $S_{ext_{B}}^{(2)}=\ln
(2TV\Delta _{0})+\frac{\chi _{0}}{\Delta _{0}}=-\frac{2}{\epsilon }-1+\ln (%
\frac{m_{0}^{4}TV}{4\pi ^{2}\epsilon })+O(\epsilon )$, where $\Delta _{0}$
and $\chi _{0}$ are defined in eq.(\ref{z2}) and eq.(\ref{z5.1}). By
comparing with eq.(\ref{z6.1}) for the particular case of identical masses
for the fermion and boson excitations we obtain%
\begin{equation*}
S_{ext_{F}}^{(2)}-S_{ext_{B}}^{(2)}=-\frac{1}{\epsilon }-\frac{5}{6}%
+O(\epsilon )
\end{equation*}%
By negleting the $1/\epsilon $ divergent term, $S_{ext_{B}}^{(2)}=\frac{5}{6}%
+S_{ext_{F}}^{(2)}$ the boson field entropy is larger than the fermion field
entropy of propagation in the space-time for identical masses.

\subsection{Free photon field entropy}

In the case of an external photon propagating, the two external points, at
zero order in $e$ reads%
\begin{equation}
\rho _{ext_{P}}^{(2,0)}=I_{\mu _{1}\mu _{2}}\int \frac{d^{4}p_{E}}{(2\pi
)^{4}}\frac{1}{p_{E}^{2}+m_{\gamma }^{2}}\left\vert p_{E}\right\rangle
\left\langle p_{E}\right\vert  \label{pho1}
\end{equation}%
where $m_{\gamma }$ is ficticious photon mass to avoid infrared divergences.
The trace reads%
\begin{equation}
Tr(\rho _{\mu \nu }^{(2,0)})=I_{\mu _{1}\mu _{2}}2TV\int \frac{d^{d}\mathbf{p%
}_{E}}{(2\pi )^{d}}\frac{1}{p_{E}^{2}+m_{\gamma }^{2}}=I_{\mu _{1}\mu
_{2}}2TV\Delta _{0}(m_{\gamma })  \label{pho2}
\end{equation}%
The quantum entropy of free photons reads%
\begin{equation}
S_{ext_{P}}^{(2,0)}=\ln (2TV\Delta _{0}(m_{\gamma }))+\frac{\chi
_{0}(m_{\gamma })}{\Delta _{0}(m_{\gamma })}  \label{pho3}
\end{equation}%
The result obtained is identical to the quantum entropy of a free scalar
boson but with $m_{0}$ replaced by $m_{\gamma }$ and the limit $m_{\gamma
}\rightarrow 0$ must be taken. From last equation an infrared divergence
appears. Nevertheless, it is well known from the theorem due to
Kinoshita-Lee and Nauenberg ( \cite{kino} and \cite{lee}) that any
physically observable must be infrared safe. To avoid the fictitious mass $%
m_{\gamma }$, a sum over additional photon emissions must be computed. This
point is very important, because in order to obtain finite values of the
observables in the infrared limit, we must consider that in the scattering
process there are some soft photons unobserved. In \cite{jsa} a mathematical
structure for this unobserved propagation was introduced. In fact, the
perturbation expansion of any quantum field theory allows rewriting the
different contributions as partial traces over some degrees of freedom that
represent particles that are not detected. In several texts, the discussion
is introduced in the context of the vertex correction to the electron
propagator. The first virtual contribution comes from a photon connecting
two electron propagator. To this virtual contribution we must add the real
soft photon contribution, that is nothing more than "opening" the virtual
photon propagator (see page 199 and page 203 of \cite{PS}). Is interesting
to note that we can avoid infrared divergences by considering that
unobserved photons are contributing. In \cite{jsa} a discussion about $\phi
^{4}$ theory implies that the first contribution to the scalar boson
propagator implies not measuring a third scalar propagator. This unobserved
boson implies tracing over its degrees of freedom and this corresponds to
"close" the propagator and obtain the loop, which introduces an ultraviolet
divergence.\footnote{%
Perhaps it could be possible to renormalize the theory by considering that
there are unobserved heavy bosons propagating anywhere that are not
measured. These heavy bosons are the equivalent to the soft unobserved
photons. These soft photons are real photons with energy less than some
cutoff $E_{c}\,$, where $E_{c}$ is the maximum photon energy allowed to
escape detection. In the same way, the heavy bosons are integrated from $%
E_{b}$ to $\infty $, and $E_{b}$ is the minimum boson energy allowed to
escape detection.} Following the same procedure, it is possible to introduce
soft photon emissions in the quantum entropy by simply adding to eq.(\ref%
{pho1}) a quantum state with ficticious mass $m_{\gamma }$ but that is
integrated in momentum from $0$ to $E_{c}\,$, where $E_{c}$ is the maximum
photon energy allowed to escape detection. Computing the eq.(\ref{pho3}) and
considering the $d\rightarrow 4$ limit, the quantum entropy of a free
photonic field reads%
\begin{equation}
S_{ext_{P}}^{(2,0)}=-\frac{2}{\epsilon }-1+\ln (\frac{E_{c}^{4}TV}{4\pi
^{2}\epsilon })+O(\epsilon )  \label{pho4}
\end{equation}%
The logarithmic behavior is identical to the free bosonic and fermionic
quantum entropies and is an universal feature of the entanglement entropy
for free quantum fields. It\ diverges with the cutoff as $\epsilon ^{-1}$
and $\ln (\epsilon )$ and the finite part depends on some complex number and
the logarithm of some dimensionless number $m_{0}^{4}TV$.\footnote{%
We are using $\hbar =c=1$, which implies that $[$energy$]=[$mass$]=[$distance%
$]^{-1}$.}

\subsection{First correction to the fermion field entropy}

In this case the total quantum operator at second order in $e$ for the
electron propagator reads 
\begin{gather}
\rho _{ext_{F}}^{(2,1)}=\int \frac{d^{d}p}{(2\pi )^{d}}\frac{d^{d}q}{(2\pi
)^{d}}\frac{i(\slashed{p}+m_{0})e^{-i\mathbf{p}\cdot \mathbf{x}_{1}}}{%
p^{2}-m_{0}^{2}}\gamma ^{\mu _{1}}\frac{i(\slashed{q}+m_{0})}{q^{2}-m_{0}^{2}%
}\gamma ^{\mu _{2}}\frac{i(\slashed{p}+m_{0})e^{i\mathbf{p}\cdot \mathbf{x}%
_{2}}}{p^{2}-m_{0}^{2}}\frac{-ig_{\mu _{1}\mu _{2}}}{(p-q)^{2}}\left\vert
x_{1}\right\rangle \left\langle x_{2}\right\vert  \label{fi1} \\
d^{4}y_{1}d^{4}y_{2}d^{4}x_{1}d^{4}x_{2}  \notag
\end{gather}%
\begin{figure}[tbp]
\centering\includegraphics[width=85mm,height=30mm]{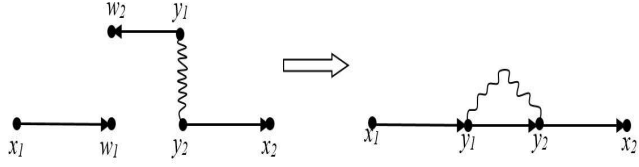}
\caption{Left: open Feynmann diagram representing the quantum operator at
second order in $e$ for the fermion propagator. Right: Partial trace over
the internal degrees of freedom $y_{1}$, $w_{1}$, $y_{2}$ and $w_{2}$ which
gives the self-energy contribution to the electron propagator.}
\label{selfel}
\end{figure}
This quantum state can be obtained by computing the trace over the internal
degrees of freedom represented by the basis $\left\vert
y_{1},y_{2}\right\rangle \left\langle w_{1},w_{2}\right\vert $ over the
quantum operator defined as (see figure \ref{selfel})%
\begin{gather}
\rho _{F}^{(2,1)}=\int S_{F}(x_{1}-y_{1})\gamma ^{\mu
_{1}}S_{F}(y_{2}-w_{1})\gamma ^{\mu
_{2}}S_{F}(w_{2}-x_{2})D_{P}(y_{1}-y_{2})g_{\mu _{1}\mu _{2}}\left\vert
x_{1},y_{1},y_{2}\right\rangle \left\langle x_{2},w_{1},w_{2}\right\vert
\label{fi1.1} \\
d^{d}x_{1}d^{d}x_{2}d^{d}y_{1}d^{d}w_{1}d^{d}y_{2}d^{d}w_{2}  \notag
\end{gather}%
where $S_{F}$ is the fermionic propagtor and $D_{P}$ is the photon
propagator. By applying Wick rotation and computing the Fourier transform,
the quantum operator reads%
\begin{gather}
Tr_{int}[\rho _{F}^{(2,1)}]=\rho _{ext_{F}}^{(2,1)}=  \label{fi1.2} \\
=-\int \frac{d^{d}p_{E}}{(2\pi )^{d}}\frac{(\slashed{p}_{E}+m_{0})}{%
p_{E}^{2}+m_{0}^{2}}\Sigma _{2}(\slashed{p}_{E})\frac{(\slashed{p}_{E}+m_{0})%
}{p_{E}^{2}+m_{0}^{2}}\left\vert p_{E}\right\rangle \left\langle
p_{E}\right\vert  \notag
\end{gather}%
where $\Sigma _{2}(\slashed{p}_{E})$ is the second order in $e$ contribution
to the self-energy (see eq.(7.16) of \cite{PS}).\footnote{%
The $e$ dependence in $\Sigma _{2}(\slashed{p}_{E})$ is considered in the
expansion of the quantum entropy of eq.(\ref{a5}).} From eq.(10.41) of \cite%
{PS}, $\Sigma _{2}$ can be written as $\Sigma _{2}(\slashed{p}_{E})=\Sigma
_{2}^{(0)}(p_{E})-\Sigma _{2}^{(1)}(p_{E})\slashed{p}_{E}$, where%
\begin{eqnarray}
\Sigma _{2}^{(0)}(p_{E}) &=&\frac{e^{2}}{(4\pi )^{d/2}}\int_{0}^{1}dx\frac{%
\Gamma (2-\frac{d}{2})(4-\epsilon )m_{0}}{[(1-x)m_{0}^{2}+x\mu
^{2}-x(1-x)p_{E}^{2}]^{2-\frac{d}{2}}}  \label{fa1} \\
\Sigma _{2}^{(1)}(p_{E}) &=&\frac{e^{2}}{(4\pi )^{d/2}}\int_{0}^{1}dx\frac{%
\Gamma (2-\frac{d}{2})(2-\epsilon )x}{[(1-x)m_{0}^{2}+x\mu
^{2}-x(1-x)p_{E}^{2}]^{2-\frac{d}{2}}}  \notag
\end{eqnarray}%
We can write $(\slashed{p}_{E}+m_{0})\Sigma _{2}(\slashed{p}_{E})(\slashed{p}%
_{E}+m_{0})=A_{0}(p_{E})+A_{1}(p_{E})\slashed{p}_{E}$, where%
\begin{eqnarray}
A_{0}(p_{E}) &=&\Sigma _{2}^{(0)}\left( p_{E}^{2}+m_{0}^{2}\right)
-2m_{0}\Sigma _{2}^{(1)}p_{E}^{2}  \label{fa2} \\
A_{1}(p_{E}) &=&m_{0}\Sigma _{2}^{(0)}-\Sigma _{2}^{(1)}\left(
m_{0}p_{E}^{2}+m_{0}^{2}\right)  \notag
\end{eqnarray}%
In order to compute $Tr[\rho _{ext_{F}}^{(2,1)}\ln (\rho
_{ext_{F}}^{(2,0)})] $, we note that the $\ln (\rho _{ext_{F}}^{(2,0)})$ has
been computed in eq.(\ref{z3.1}), so that%
\begin{gather}
Tr[\rho _{ext_{F}}^{(2,1)}\ln (\rho _{ext_{F}}^{(2,0)})]=-2TV\int \frac{%
d^{d}p_{E}}{(2\pi )^{d}}\frac{A_{0}(p_{E})\ln (p_{E}^{2}+m_{0}^{2})}{%
2(p_{E}^{2}+m_{0}^{2})^{2}}  \label{fa3} \\
-2TV\int \frac{d^{d}p_{E}}{(2\pi )^{d}}\frac{A_{1}(p_{E})p_{E}}{%
2(p_{E}^{2}+m_{0}^{2})^{2}}\ln (\frac{m_{0}+p_{E}}{m_{0}-p_{E}})  \notag
\end{gather}%
where we have neglected the odd terms in $\slashed{p}_{E}$ because they
integrate symmetrically to zero. In turn, the trace of $\rho
_{ext_{F}}^{(2,1)}$ reads%
\begin{equation}
\beta ^{(2,1)}=-2TV\int \frac{d^{d}p_{E}}{(2\pi )^{d}}\frac{A_{0}(p_{E})}{%
(p_{E}^{2}+m_{0}^{2})^{2}}  \label{fa4}
\end{equation}%
Eqs.(\ref{fa4}) and (\ref{fa3}) are complicated integrals that give the
second order contribution to the fermion entropy. Instead of computing the
last integrals, we can consider a more simple system in which the full
propagator can be solved exactly. This model is the Bloch-Nordsieck model 
\cite{Bloch}, where the Dirac matrices $\gamma ^{\mu }$ in the Lagrangian
are replaced by $u^{\mu }$, where $u^{\mu }$ are the components of a
velocity vector and $u^{\mu }u_{\mu }=1$. This model has been solved in \cite%
{tarski} and an exact solution to the full Green function reads (see \cite%
{bogo}, eq.(46.28), page 484) 
\begin{equation}
G(p)=\frac{1}{(u_{\mu }p^{\mu }-m_{0})^{\gamma +1}}  \label{fa5}
\end{equation}%
\begin{figure}[tbh]
\begin{minipage}{0.48\linewidth}
\includegraphics[width=70mm,height=40mm]{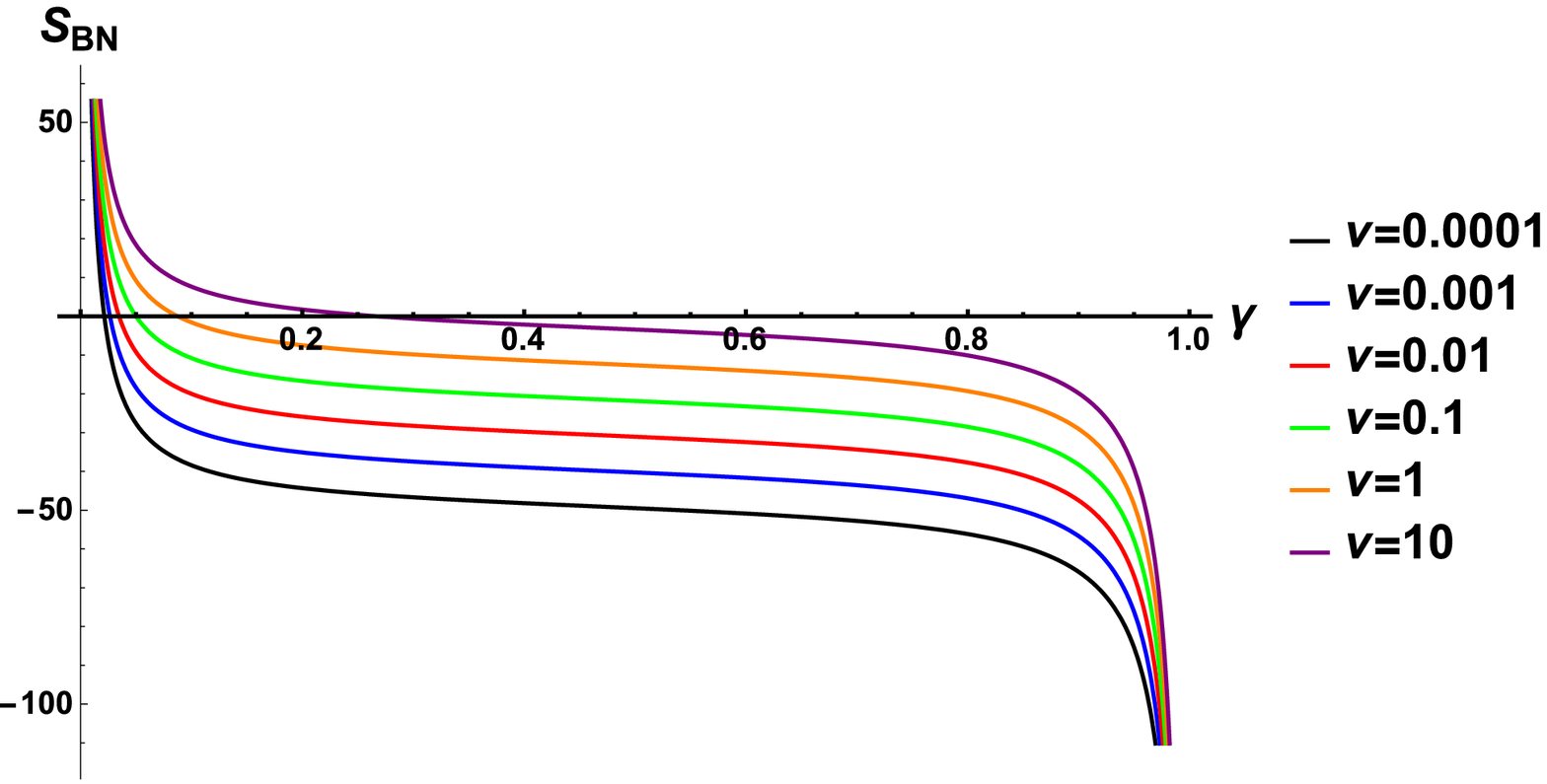} 
\label{bloch}
\end{minipage}
\hspace{0.08cm} 
\begin{minipage}{0.5\linewidth}
\centering
\includegraphics[width=70mm,height=40mm]{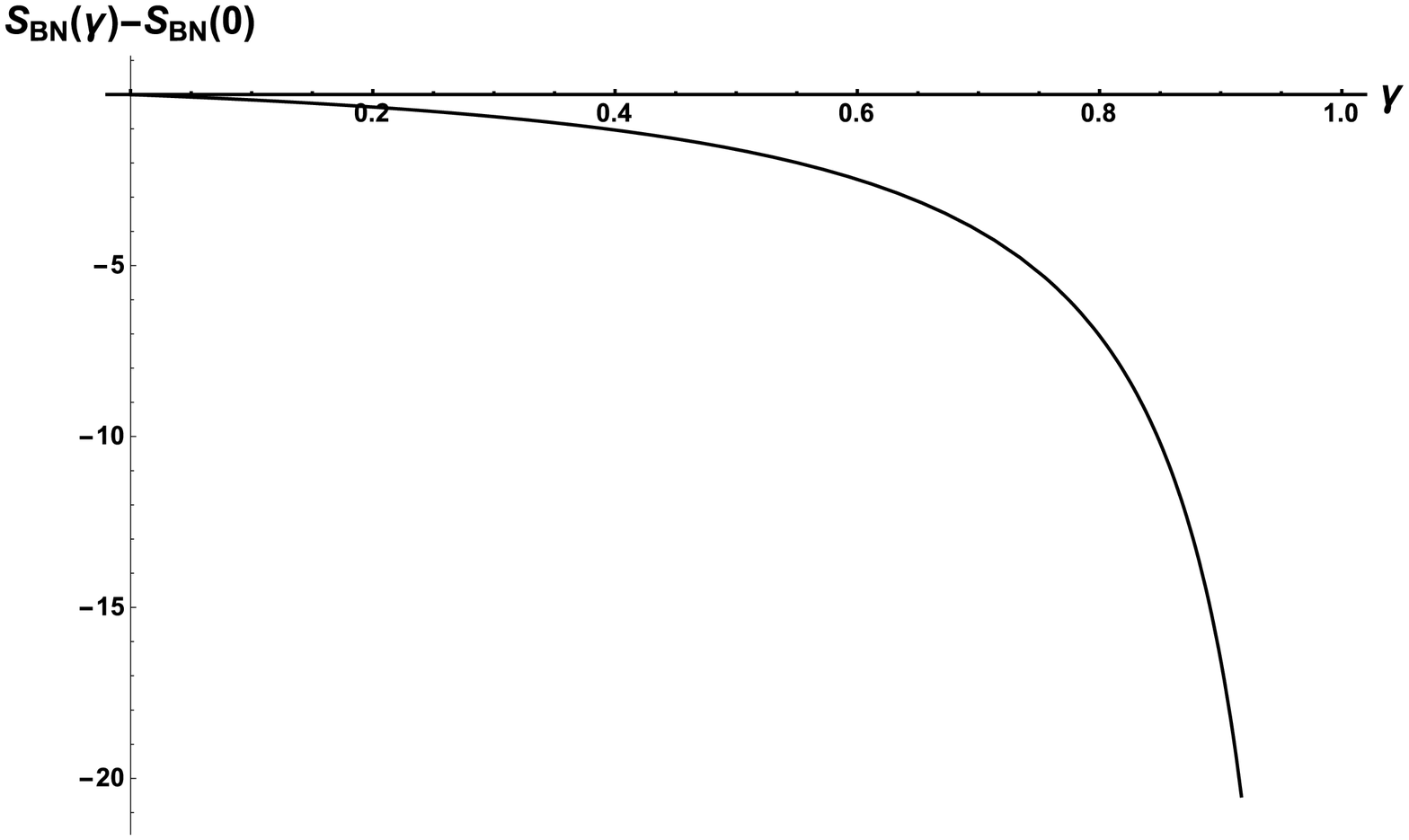}
\label{bloch2}
\end{minipage}
\caption{Right: Total entropy as a function of $\protect\gamma $ for
different values of the ratio $\protect\nu =\frac{m_{0}^{4}TV}{\cos ^{4}%
\protect\theta }$ for the Bloch-Nordsieck model. Left: Difference between
total entropy and entropy without interactions for the Bloch-Nordsieck model
as a function of $\protect\gamma =\frac{\protect\alpha }{2\protect\pi }(3-%
\protect\xi )$. }
\label{bloch}
\end{figure}
where $\gamma =\frac{e^{2}}{8\pi ^{2}}(3-\xi )=\frac{\alpha }{2\pi }(3-\xi )$%
, where $\alpha $ is the fine structure constant and $\xi $ is a gauge
fixing parameter. We can write $u_{\mu }p^{\mu }=\left\vert p\right\vert
\cos \theta $, where $\theta $ is the angle between $u$ and $p$. This full
propagator is the analogue to the full propagator of a $\phi ^{4}$ theory
written in terms of the partial trace of a quantum density operator (see 
\cite{PR1}, eq.(65)) or the full electron propagator of QED, $%
G=(p^{2}-m_{0}^{2}-\Sigma (\slashed{p}))^{-1}$. As we write the quantum
operator for the electron or boson propagator, we can do the same with the
quantum state in the Bloch-Nordsieck model as%
\begin{equation}
\rho =\int \frac{d^{4}p}{(2\pi )^{4}}\frac{1}{(p\cos \theta -m_{0})^{\gamma
+1}}\left\vert p\right\rangle \left\langle p\right\vert  \label{fa6}
\end{equation}%
The quantum entropy can be computed as $S_{BN}=\ln \left[ 2TV\Delta \right] +%
\frac{(\gamma +1)}{\Delta }\Gamma $, where%
\begin{equation}
\Delta =\frac{1}{8\pi ^{2}}\int_{0}^{\infty }\frac{p^{3}}{(p\cos \theta
-m_{0})^{\gamma +1}}dp=\frac{6(-1)^{3-\gamma }(m_{0})^{3-\gamma }}{8\pi
^{2}(\gamma -3)(\gamma -2)(\gamma -1)\gamma \cos ^{4}\theta }  \label{fa7}
\end{equation}%
and%
\begin{gather}
\Gamma =\frac{1}{8\pi ^{2}}\int \frac{p^{3}\ln (up-m_{0})}{%
(up-m_{0})^{\gamma +1}}dp=  \label{fa8} \\
\frac{6(-1)^{\gamma }(m_{0})^{\gamma }m_{0}^{3}[-2(2\gamma -3)(1+\gamma
(\gamma -3))-(\gamma -3)(\gamma -2)\gamma \ln (-m_{0})]}{8\pi ^{2}(\gamma
-3)^{2}(\gamma -2)^{2}(\gamma -1)^{2}\gamma ^{2}\cos ^{4}\theta }  \notag
\end{gather}%
Then%
\begin{equation}
S_{BN}=\frac{1+\gamma }{\gamma }+\frac{1+\gamma }{\gamma -1}+\frac{1+\gamma 
}{\gamma -2}+\frac{1+\gamma }{\gamma -3}+\ln [\frac{3m_{0}^{4}TV}{2\pi
^{2}\cos ^{4}\theta (\gamma -3)(\gamma -2)(\gamma -1)\gamma }]  \label{fa9}
\end{equation}%
By considering the limit $\gamma \rightarrow 0$, $\Delta
S_{BN}=S_{BN}(\gamma )-S_{BN}(0)$ reads%
\begin{equation}
\Delta S_{BN}=\frac{11}{6}+\frac{1+\gamma }{\gamma -1}+\frac{1+\gamma }{%
\gamma -2}+\frac{1+\gamma }{\gamma -3}+\ln [\frac{6}{(3-\gamma )(2-\gamma
)(1-\gamma )}]  \label{fa9.1}
\end{equation}%
In figure \ref{bloch} the total entropy for different values of $\frac{%
m_{0}^{4}TV}{\cos ^{4}\theta }$ is shown in the first case and the
difference of the total entropy with respect the non-interacting case $%
\Delta S_{BN}$ is shown as a function of $\gamma $ in the second case. As it
can be seen, the interactions decrease the fermion entropy. In fact, by
replacing $\gamma $ by $\frac{\alpha }{\pi }$, where the Feynman gauge is
considered $\xi =1$, we obtain $\Delta S_{BN}\sim -0.003$, which is the
entropy lost by the interactions. This is the same behaviour found in the
quantum entropy of the boson field. In \cite{jsa}, it was shown that the
quantum entropy at first order in $\lambda _{0}$ for the boson field reads%
\begin{gather}
S_{B}^{ext}=\ln (2TV\Delta _{0})+\frac{\chi _{0}}{\Delta _{0}}+\frac{\lambda
_{0}\mu ^{-\epsilon }}{2}\left( \chi _{1}-\frac{\Delta _{1}\chi _{0}}{\Delta
_{0}}\right) +O(\lambda _{0}^{2})  \label{fa9.2} \\
=-\frac{2}{\epsilon }-1+\ln (\frac{m_{0}^{4}TV}{4\pi ^{2}\epsilon })+\frac{%
\lambda _{0}}{32\pi ^{2}}\left( -1+2\gamma _{0}+\ln (\frac{m_{0}^{4}}{4\pi
^{2}\mu ^{4}})\right) +O(\lambda _{0}^{2})  \notag
\end{gather}%
where $\gamma _{0}$ is the Euler-Mascheroni constant. The contribution at
order $\lambda _{0}$ is similar to the results obtained in \cite{bala} for
the mutual information. In figure \ref{boson}, the total entropy is plotted
as a function of $\frac{m_{0}}{\mu }$, where it can be seen that the
contribution at first order in $\lambda _{0}$ decreases the quantum entropy
with respect the free value. This result for the Bloch-Nordsieck and the
scalar boson suggests that interactions reduce the unpredictability of the
quantum operator propagation. 
\begin{figure}[tbp]
\centering\includegraphics[width=100mm,height=60mm]{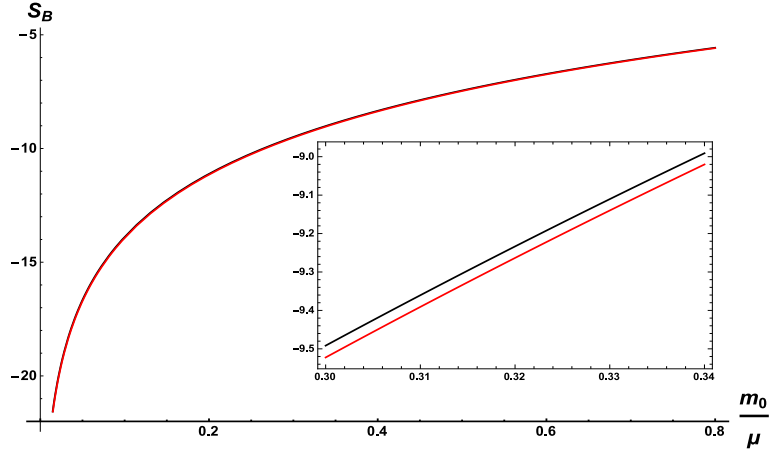}
\caption{Quantum entropy of boson propagator at zero order in $\protect%
\lambda _{0}$ (black line) and at first order in $\protect\lambda _{0}$ (red
line). }
\label{boson}
\end{figure}
On the other hand, by computing the integrals of eq.(\ref{fa7}) and eq.(\ref%
{fa8}) in $d$ dimensions, taking the limit $d\rightarrow 4$ and finally the $%
\gamma \rightarrow 0$ limit, the quantum entropy can be written as%
\begin{equation}
S_{BN}=-\frac{1}{\epsilon }-\frac{11}{6}+\ln (\frac{m_{0}^{4}TV}{4\pi
^{2}\epsilon \cos ^{4}\theta })-\frac{49}{36}\gamma -\frac{199}{108}\gamma
^{2}-O(\gamma ^{3})  \label{fa10}
\end{equation}%
where the finite term $-\frac{11}{6}$ is identical to the QED interaction
(see eq.(\ref{z6.1})). Without loss of generality, taking $\cos \theta =1$,
then the free quantum entropy obtained follows the same behavior as the
quantum entropy for free fermions (see eq.(\ref{z6.1}). The logarithm term $%
\ln (\frac{m_{0}^{4}TV}{4\pi ^{2}\epsilon })$ is universal for the different
quantum fields.

\subsection{First correction to the photon field entropy}

\begin{figure}[tbp]
\centering\includegraphics[width=80mm,height=30mm]{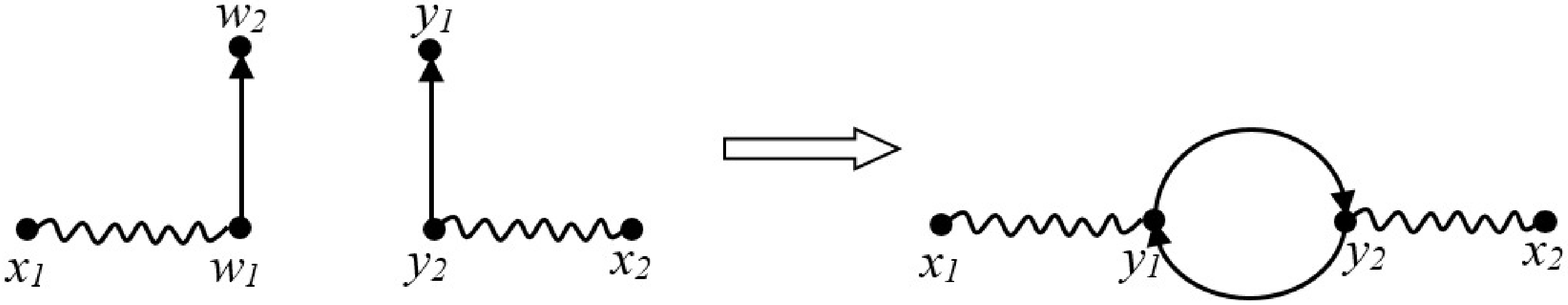}
\caption{Left: open Feynmann diagram representing the quantum operator at
second order in $e$ for the photon propagator. Right: Partial trace over the
internal degrees of freedom $y_{1}$, $w_{1}$, $y_{2}$ and $w_{2}$ which
gives the self-energy contribution to the photon propagator. }
\label{selfph}
\end{figure}

The quantum operator of the first correction to the photon propagator reads
(see figure \ref{selfph}) 
\begin{gather}
\rho _{P}^{(2,1)}=\int D_{P}(x_{1}-y_{1})g_{\mu _{1}\rho
}S_{F}(y_{1}-w_{2})\gamma ^{\rho }\gamma ^{\nu
}S_{F}(w_{1}-y_{2})D_{P}(y_{2}-x_{2})g_{\nu \mu _{2}}\times  \label{fot1} \\
\left\vert x_{1},y_{1},y_{2}\right\rangle \left\langle
x_{2},w_{1},w_{2}\right\vert
d^{d}x_{1}d^{d}x_{2}d^{d}y_{1}d^{d}w_{1}d^{d}y_{2}d^{d}w_{2}  \notag
\end{gather}%
The partial trace over the internal degrees of freedom $y_{1}$, $\omega _{1}$
and $y_{2}$, $w_{2}$ gives as a result the first quantum correction to the
photon propagator reads%
\begin{equation}
\rho _{ext_{P}}^{(2,1)}=\int \frac{d^{d}p}{(2\pi )^{d}}\frac{\Pi _{2}^{\mu
_{1}\mu _{2}}(p)}{(p^{2}+\mu ^{2})^{2}}\left\vert p\right\rangle
\left\langle p\right\vert  \label{fot2}
\end{equation}%
where we have used that introduced the Fourier transform of $\left\vert
x_{1}\right\rangle $ and $\left\langle x_{2}\right\vert $ and where (see
eq.(7.71) of \cite{PS})%
\begin{equation}
\Pi _{2}^{\mu _{1}\mu _{2}}(p)=\int \frac{d^{d}q}{(2\pi )^{d}}tr\left[
\gamma ^{\mu _{1}}\frac{i}{\slashed{q}-m_{0}}\gamma ^{\mu _{2}}\frac{i}{%
\slashed{q}+\slashed{p}-m_{0}}\right]  \label{fot3}
\end{equation}%
which in turn can be written as $\Pi _{2}^{\mu _{1}\mu _{2}}(p)=(p^{2}\delta
_{\mu _{1}\mu _{2}}-p^{\mu _{1}}p^{\mu _{2}})\Pi _{2}(p)$, where%
\begin{equation}
\Pi _{2}(p)=-\frac{8}{(4\pi )^{d/2}}\int_{0}^{1}dx\frac{x(1-x)\Gamma (2-%
\frac{d}{2})}{[m_{0}^{2}-x(1-x)p^{2}]^{2-\frac{d}{2}}}  \label{fot4}
\end{equation}%
The quantum entropy at second order in $e$ implies to compute two integrals%
\begin{equation}
\beta ^{(2,1)}=Tr[\rho _{ext_{P}}^{(2,1)}]=2TV\int \frac{d^{d}p}{(2\pi )^{d}}%
\frac{\Pi _{2}^{\mu _{1}\mu _{2}}(p)}{(p^{2}+\mu ^{2})^{2}}  \label{fot5}
\end{equation}%
and%
\begin{equation}
Tr[\rho _{ext_{P}}^{(2,1)}\ln (\rho _{ext_{P}}^{(2,1)})]=-2TV\int \frac{%
d^{d}p}{(2\pi )^{d}}\frac{\Pi _{2}^{\mu _{1}\mu _{2}}(p)\ln (p^{2}+\mu ^{2})%
}{(p^{2}+\mu ^{2})^{2}}  \label{fot6}
\end{equation}%
where we have disregarded terms with odd $p_{\mu _{E}}$ and $q_{\mu _{E}}$
in the numerator. Last integrals can be solved in order to obtain the first
contribution to the quantum entropy of the photon propagation. Considering
the Bloch-Nordsieck model, in contrast with the fermionic self-energy, there
is no vacuum polarization, which is the effect of the photon self-energy.
Then it is not possible to obtain other contributions to the quantum entropy
in this model than the result obtained in eq.(\ref{pho4}).

Summing up, we can collect all the results for the free quantum entropies
for different quantum fields%
\begin{gather}
S_{B}=-\frac{2}{\epsilon }-1+\ln (\frac{m_{B}^{4}TV}{4\pi ^{2}\epsilon })
\label{su1} \\
S_{F}=-\frac{1}{\epsilon }-\frac{11}{6}+\ln (\frac{m_{F}^{4}TV}{4\pi
^{2}\epsilon })  \notag \\
S_{P}=-\frac{2}{\epsilon }-1+\ln (\frac{E_{c}^{4}TV}{4\pi ^{2}\epsilon }) 
\notag
\end{gather}%
which can be condensed in 
\begin{equation}
S_{i}=-\frac{a_{i}}{\epsilon }-b_{i}+\ln (\frac{m_{i}^{4}TV}{4\pi
^{2}\epsilon })  \label{su2}
\end{equation}%
where we can consider that the maximum photon energy allowed to escape
detection $E_{l}$ can be considered as an of-shell photon mass. For any two
scalar bosons with different masses $m_{B_{1}}$ and $m_{B_{2}}$we have that $%
S_{B_{1}}-S_{B_{2}}=4\ln (\frac{m_{B_{1}}}{m_{B_{2}}})$. In turn, $%
S_{B}-S_{F}=\frac{5}{6}+4\ln (\frac{m_{B}}{m_{F}})$ and if $m_{B}>e^{-\frac{5%
}{24}}m_{F}$ then $S_{B}>S_{F}$. The different quantum entropies contain
ultraviolet divergences which can be isolated by dimensional regularization.
It should be stressed that even in the most simple case where no
interactions are considered, the von Neumann entropy contains ultraviolet
divergences (see eq.(\ref{su1})). This implies that no mathematical
operation at the level of the density quantum operators exists to avoid UV
divergences. The von Neumann entropy for the free scalar propagator depends
only on the mass of the quantum field $m_{0}$, the space-time volume and the
ultraviolet cutoff. These divergences appear similarly in the entanglement
entropy between regions of space-time \cite{casini}. In local quantum field
theory, discussions of entanglement are focused on the density matrices
associated with bounded spatial regions. These results are well-defined
because by locality, there are independent degrees of freedom in disjoint
spatial domains, so the Hilbert space factorizes. The associated spatial
entanglement entropy is typically divergent, even in free field theory,
because in the continuum limit, any spatial region contains an infinite
number of degrees of freedom produced by high energy vacuum fluctuations at
arbitrarily short wavelengths. These divergences require regularization and
some procedure is needed to extract finite regularization independent data.

In entanglement entropy between space-time regions, the terms that are
proportional to $\frac{1}{\epsilon ^{j}}$ are not physical since they are
not related to quantities well defined in the continuum (\cite{casini}). The
logarithmic divergence is expected to be universal in the sense that is
independent of the regularization prescription adopted or of the microscopic
model used to obtain the continuum QFT at distances large with respect to
the cutoff.\footnote{%
Perhaps these similar terms imply a deep connection between entanglement
between space-time regions and local interactions between fields. In turn,
if this deep connection turns to be an identity, then model introduced in
this manuscript can be useful to compute entanglement entropy between curved
space-time regions.}

\section{Conclusions}

In this work, the entanglement entropy between real and virtual propagating
states has been computed by rewriting the generating functional of the
quantum electrodynamics theory in terms of quantum operators and inner
products. In this way, it is possible to compute the von Neumann entropy for
the electron and photon propagator as a perturbation expansion in $e$. It
was shown that for the Bloch-Nordsieck model, the interactions decrease the
quantum entropy with respect the non-interacting case. In turn, it is shown
the universal behavior of the von Neumann entropy for different free quantum
fields, that depends on the logarithm of the dimensionless parameter $\frac{%
m^{4}TV}{\epsilon }$ and some particular constants. The first order
contributions to the entropy of the fermion and photon fields are considered
and the results are computed in terms of complex integrals. The formalism
introduced can be useful to characterize the entanglement entropy that
interactions introduce. In turn, the entanglement can be understood as
unobserved field excitations which are traced out.

\section{Acknowledgment}

This paper was partially supported by grants of CONICET (Argentina National
Research Council) and Universidad Nacional del Sur (UNS) and by ANPCyT
through PICT 1770, and PIP-CONICET Nos. 114-200901-00272 and
114-200901-00068 research grants. J. S. A. is member of CONICET.

\appendix

\section{Appendix A}

In order to get closer the ideas of this manuscript and the general boundary
formalism only for boundaries defined by spacelike hyperplanes consider the
quantum scalar field%
\begin{equation}
\phi _{0}(x)=\int \frac{d^{3}p}{(2\pi )^{3}}\frac{1}{\sqrt{2E_{p}}}\left( 
\mathbf{a}_{p}e^{ipx}+\mathbf{a}_{p}^{\dag }e^{-ipx}\right)  \label{r1}
\end{equation}%
then consider this quantum field as the coordinate representation of a ket $%
\left\vert \phi _{0}(x_{1})\right\rangle $ in the space-time coordinate $%
x_{1}$ and another quantum field in the space-time coordinate $x_{2}$, that
is $\left\vert \phi _{0}(x_{2})\right\rangle $, where the time component of $%
x_{1}$ is smaller than the time component of $x_{2}$ (see \cite{oeckl3}
below eq.(3)). If we suppose that the time-component of $x_{1}$ is smaller
than the time component of $x_{2}$ and the space coordinates can vary over a
space-like hyperplane, then we can define the quantum density operator 
\begin{equation}
\rho _{0}=\left\vert \phi _{0}(x_{1})\right\rangle \left\langle \phi
_{0}(x_{2})\right\vert  \label{r2}
\end{equation}%
then is not difficult to show that%
\begin{equation}
\left\langle \Omega _{0}\left\vert \rho _{0}\right\vert \Omega
_{0}\right\rangle =\Delta _{0}(x_{1}-x_{2})  \label{r3}
\end{equation}%
that is, the coefficient of the quantum operator of eq.(31)\ of \cite{jsa}
is the vacuum expectation value of the quantum density operator defined in
eq.(\ref{r2}), that is $\rho =\int \left\langle \Omega _{0}\left\vert \rho
_{0}\right\vert \Omega _{0}\right\rangle \left\vert x_{1}\right\rangle
\left\langle x_{2}\right\vert d^{3}\mathbf{r}_{1}d^{3}\mathbf{r}_{2}$, where
the time components of $x_{1}$ and $x_{2}$ are fixed and not integrated.
This quantum operator is suitable for processes where a preparation is done
in $t_{1}$ and a measurement is done in $t_{2}$ or more simpler a creation
and a later annihilation of a field excitation.\footnote{%
In turn, this quantum density operator manifest naturally the in-out
duality, which blurs the distinction between preparation and observation
proper in the measurement \cite{oeckl1} due to the interchange of in and out
coordinates. This is in turn what the LSZ reduction manifest, where the
correlation functions written in the momentum space do not depends on the
choice of incoming and outgoing momentum.} For virtual processes this
quantum state is not suitable because the perturbation expansion demands
that an integration $\int d^{4}y$ must be computed (is the superposition
principle \cite{PS}, p. 94).

For external points is not possible to restrict the quantum operator to a
single time slice because the time component of $x_{1}$ must be smaller than
the time component of $x_{2}$ and in turn $t_{1}$ and $t_{2}$ must be fixed.
For virtual propagations there is no restriction and can be the case in
which $t_{2}=t_{1}$, that is, the quantum operator is restricted to a single
time-slice. The procedure done in \cite{jsa} manifest this virtual process
as a real propagator $\Delta (y_{1}-w_{1})$ between two arbitrary space-time
coordinates $y_{1}$ and $w_{1}$ and a sum over all the possible space-time
coordinates $y_{1}$ and $w_{1}$ must be done. This sum is provided by the
lack of measurement of these two space-time points by introducing the Dirac
delta distribution $\delta (y_{1}-w_{1})$ as the internal part of the
observable. From this point of view, there is an identification of virtual
propagation with real propagation by opening the loop $y_{1}\rightarrow
y_{1} $ to $y_{1}\rightarrow w_{1}$\footnote{%
The order of the coordinates is irrelevant.}. This happens only when
interactions are turned on. Processes as propagation or scattering events
happen inside a space-time region, which is the space-time region relevant
for the experiment, in the sense that the particle inflow and detection
happens on the boundary of this space-time region. The interaction term in
the Lagrangian is turned on only inside the boundary. The particles detected
on the boundary should be considered as free. In this sense, the formalism
introduced above treats observables as located in spacetime regions and
giving rise to linear maps from the region's boundary Hilbert space to
complex numbers. The boundary Hilbert space is a tensor product of the
preparation and measurement Hilbert spaces. For no interactions, the quantum
state is not mixed, it only consists of the tensor product of the prepared
quantum state and the measured quantum state. When interactions are turned
on, the quantum state cannot be written as a tensor product, but not because
of the bulk effects on the boundary but rather by the entanglement between
the real and the virtual states. This virtual state can be translated to the
boundary, but it must remain unobserved. The lack of observation (lack of
preparation or measurement of this new state) implies to compute the partial
trace over the degrees of freedom of the total quantum density operator.

Then, a relationship between the interaction terms in the Lagrangian and the
undetermined metric of space-time in the bulk of the boundary defined by the
preparation and measurement can be done. For example, if we consider two
time-slices in flat-space time, $S=S_{1}\cup S_{2}$ with $S_{1}$ at $t_{1}$
and $S_{2}$ at $t_{2}$, the bulk is the region between the time-intervals $%
[t_{1},t_{2}]$. That is, the boundary metric is fixed and defined by the
observers, but nothing can be said about the interior of the boundary (see 
\cite{oeckl1}). In \cite{bianchi}, the distinction between pure and mixed
states is weaken in the general covariant context when finite spatial
regions are considered. In the model introduced in this paper, the quantum
state is mixed when interactions are turned on. The mixture is due to the
entanglement of the virtual state in the bulk with the real states in the
boundary. In turn, for free fields there is a priori distinction between
pure and mixture states because we can distinguish between past and future
parts of the boundary. Moreover, the observables acts in the infinite past
and infinite future. In this sense, it seems that the model introduced in
this work is a particular case of the general boundary formalism with the
incorporation of the interactions treated in a perturbative manner and
allowing these virtual states to be defined in the whole space-time.

\section{Appendix B}

To solve eq.(\ref{z3.1}) we can note that if two matrices $A$ and $B$
commute, then $\ln (AB)=\ln (A)+\ln (B)$, then 
\begin{equation}
\ln (\frac{\slashed{p}_{E}+m_{0}}{p_{E}^{2}+m_{0}^{2}})=\ln (\frac{1}{%
p_{E}^{2}+m_{0}^{2}})I+\ln (\slashed{p}_{E}+m_{0})  \label{ap1}
\end{equation}%
the second term of last equation can be written as 
\begin{equation}
\ln (\slashed{p}_{E}+m_{0})=\ln (m_{0})+\ln [\frac{\slashed{p}_{E}}{m_{0}}+I]
\label{ap2}
\end{equation}%
Using the Mercator expansion $\ln (I+K)=\overset{+\infty }{\underset{n=1}{%
\sum }}\frac{(-1)^{n+1}}{n}K^{n}$, where $K=\frac{\slashed{p}_{E}}{m_{0}}$
and using that $\slashed{p}_{E}\slashed{p}_{E}=(-i\slashed{p})^{2}=p_{E}^{2}$%
, then $K^{2}=(\frac{p_{E}}{m_{0}})^{2}$, $K^{3}=\frac{\slashed{p}_{E}}{m_{0}%
}(\frac{p_{E}}{m_{0}})^{2}$, $K^{4}=(\frac{p_{E}}{m_{0}})^{4}$, $K^{5}=\frac{%
\slashed{p}_{E}}{m_{0}}(\frac{p_{E}}{m_{0}})^{4}$, $K^{6}=(\frac{p_{E}}{m_{0}%
})^{6}$, etc, last equation can be written as\footnote{%
Must be stressed that the Mercator expansion converges to $\ln (K+I)$ for $%
\left\vert \frac{p}{m_{0}}\right\vert <1$, but an analytical continuation to
the entire complex plane can be applied.}%
\begin{equation}
\ln [\frac{\slashed{p}_{E}}{m_{0}}+I]=\frac{\slashed{p}_{E}}{m_{0}}\overset{%
+\infty }{\underset{n=1}{\sum }}\frac{1}{2n-1}(\frac{p_{E}}{m_{0}})^{2n-2}-%
\overset{+\infty }{\underset{n=1}{\sum }}\frac{1}{2n}(\frac{p_{E}}{m_{0}}%
)^{2n}  \label{ap3}
\end{equation}%
using that $\overset{+\infty }{\underset{n=1}{\sum }}\frac{1}{2n-1}x^{2n-2}=%
\frac{1}{2x}\ln (\frac{1+x}{1-x})$ and $\overset{+\infty }{\underset{n=1}{%
\sum }}\frac{1}{2n}x^{2n}=-\frac{1}{2}\ln (1-x^{2})$, last equation read%
\begin{equation}
\ln [\frac{\slashed{p}_{E}}{m_{0}}+I]=\frac{\slashed{p}_{E}}{2p_{E}}\ln (%
\frac{m_{0}+p_{E}}{m_{0}-p_{E}})+\frac{1}{2}\ln (\frac{m_{0}^{2}+p_{E}^{2}}{%
m_{0}^{2}})  \label{ap4}
\end{equation}%
Collecting all the terms from eq.(\ref{ap1}) we obtain%
\begin{equation}
\ln (\frac{\slashed{p}_{E}+m_{0}}{p_{E}^{2}+m_{0}^{2}})=-\frac{1}{2}\ln
(p_{E}^{2}+m_{0}^{2})I+\frac{\slashed{p}_{E}}{2p_{E}}\ln (\frac{m_{0}+p_{E}}{%
m_{0}-p_{E}})  \label{ap5}
\end{equation}%
This result will be used in Section II.

\end{document}